\documentstyle[twoside,fleqn,espcrc2,epsf]{article}

\newcommand{\be}{\begin{equation}}
\newcommand{\ee}{\end{equation}}
\newcommand{\bdm}{\begin{displaymath}}
\newcommand{\edm}{\end{displaymath}}
\newcommand{\<}{\langle}
\renewcommand{\>}{\rangle}

\def\ham{{\bf \rm H}}

\def\bmat{{\bf \rm B}}
\def\cmat{{\bf \rm C}}

\def\spose#1{\hbox to 0pt{#1\hss}}
\def\ltapprox{\mathrel{\spose{\lower 3pt\hbox{$\mathchar"218$}}
 \raise 2.0pt\hbox{$\mathchar"13C$}}}
\def\gtapprox{\mathrel{\spose{\lower 3pt\hbox{$\mathchar"218$}}
 \raise 2.0pt\hbox{$\mathchar"13E$}}}
\def\inapprox{\mathrel{\spose{\lower 3pt\hbox{$\mathchar"218$}}
 \raise 2.0pt\hbox{$\mathchar"232$}}}

\title{ 
Evidence for fractional topological charge in SU(2)
pure Yang-Mills theory\thanks{Presented by U.~M.~Heller at {\sl Lattice '98.}}
}

\author{
Robert G.~Edwards, Urs M.~Heller and Rajamani Narayanan
\address{
SCRI, The Florida State University, 
Tallahassee, FL 32306-4130, USA}
}

\begin{document}

\begin{abstract}
We investigate the spectral flows of the hermitian Wilson-Dirac
operator in the fundamental and adjoint representations on two
ensembles of pure SU(2) gauge field configurations at the same
physical volume.  We find several background gauge field
configurations where the index of the hermitian Wilson-Dirac operator
in the adjoint representation is not four times the index in the
fundamental representation. This could imply a topological basis for
the existence of degenerate vacua in supersymmetric Yang-Mills theories.

\end{abstract}

\maketitle


The overlap formalism for constructing a chiral gauge theory on the
lattice~\cite{over} provides a natural definition of the index, $I$,
of the associated chiral Dirac operator. The index is equal to half
the difference of negative and positive eigenvalues of the hermitian
Wilson-Dirac operator
\be
\ham_L(m) =
\pmatrix {\bmat(U) -m & \cmat(U) \cr \cmat^\dagger(U) &  -\bmat(U) +m
\cr}
\label{eq:WDH}
\ee
-- we will use an unconventional sign for the mass term throughout! --
where $\cmat$ is the (naive) lattice transcription of the Weyl term, and
$\bmat$
is the usual Wilson term (covariant Laplacian). In the naive continuum
limit, $\ham(m) = \gamma_5 (\gamma_\mu D_\mu -m)$, $m$ can be chosen as
any positive value, since zero eigenvalues of $\ham(m)$, and therefore
eigenvalues crossing zero as function of $m$, are only possible at $m=0$.
Furthermore, crossings of zero are associated with the topology of the
background gauge field: the topological charge, $Q$, is equal to the index
of the chiral Dirac operator.

On the lattice, because of the additive mass renormalization, the
crossings of zero occur at positive $m$ and spread out in $m$. 
It is easy to see that no eigenvalues of $\ham_L(m)$ can be zero for
$m<0$. Since in the free case the first doublers become massless at $m=2$
we restrict ourselves to $m < 2$. A simple way to compute the
index $I$ is to compute the lowest eigenvalues of $\ham_L(m)$ at some
suitably small $m$ before any crossings of zero occurred. Then $m$ is
slowly varied and the number and direction of zero crossings are tracked. 
The net number at some $m_t$ is the index of the overlap chiral Dirac
operator. 

\vskip -10cm
\rightline{FSU-SCRI-98C-93}
\vskip +9.6cm

We have applied this procedure to compute the index, which we take as
the definition of topological charge on the lattice, of various gauge
field ensembles with gauge group SU(3) in \cite{su3_top}. We found
that the zero crossings start occurring at some, ensemble dependent,
$m_1>0$ and continue occurring for all $m$ in $m_1<m<2$. We found a
monotonic relation between the crossing point $m$ and the size of the
corresponding zero mode, with the crossings for larger objects occurring at
smaller $m$. All crossings for objects with size larger than about two
lattice spacings occurred within a small region of $m>m_1$. All later
zero crossings correspond to small objects of size about one to two lattice
spacings. These small objects do not seem to have physical effects and,
for example, do not affect the topological
susceptibility~\cite{su3_top,Rajamani,Robert}.

\begin{figure}
\epsfysize=3.5in
\centerline{\epsfbox[100 70 576 601]{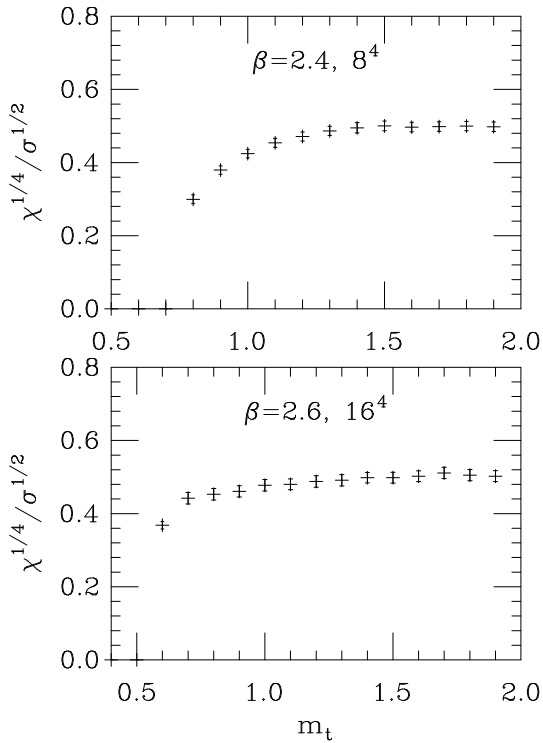}}
\caption{The topological susceptibility as function of the mass $m_t$
used to define the index of the overlap chiral Dirac operator.}
\label{fig:chi}
\end{figure}

Here we consider two ensembles of SU(2) background gauge fields
generated using the standard Wilson action with periodic boundary
conditions on the gauge fields. One is at a lattice coupling of
$\beta=2.4$ on an $8^4$ lattice and the other is at a lattice coupling
of $\beta=2.6$ on a $16^4$ lattice. The couplings were chosen so that
both lattices had the same physical volume ($a=0.12$ fm at $\beta=2.4$ and
$a=0.06$ fm at $\beta=2.6$) and are in the confined phase ($\beta_c=2.5115$
at $N_\tau=8$ and $\beta_c=2.74$ at $N_\tau=16$)~\cite{Fingberg}.
The topological susceptibility obtained from the index in the
fundamental representation is shown in Fig.~\ref{fig:chi}. As for SU(3)
we find that the topological susceptibility becomes independent of the
mass $m_t$ used to define the index, once $m_t$ is in the range of zero
crossings associated with objects of small size.

The index of the massless Dirac operator in the adjoint representation
of the SU($N$) gauge group in a background field of topological charge
$Q$ is $I_a=2NQ$~\cite{ad_index}. Classical instantons carry an
integer topological charge and therefore give a non-zero expectation
value to an operator that contains $2N$ Majorana fermions,
$\lambda(x_1)\lambda(x_1)\cdots\lambda(x_N)\lambda(x_N)$, independent
of the $x_i$ (see \cite{Shifman} for a recent review).
On very general grounds~\cite{Witten}, we know that the theory has $N$
degenerate vacua in a finite physical volume with a different non-zero
expectation value for $\lambda(x)\lambda(x)$ in each of these vacua,
and hence a $\lambda(x)\lambda(x)$ condensate in the infinite volume limit.
Self-dual twisted gauge field configurations with a fractional
topological charge of $1\over N$~\cite{Hooft} can be used to explain 
these N degenerate vacua~\cite{Cohen}. A non-perturbative computation of 
$\< \lambda(x)\lambda(x) \>$ on the lattice would typically be done with
conventional periodic boundary conditions for both the gauge fields and
fermions in order to achieve a supersymmetric theory in the continuum limit. 
In this situation the picture of N degenerate vacua in a finite physical
volume does not clearly emerge.

To shed more light on this question, we use the overlap definition of the
index of the chiral Dirac operator in the adjoint representation, $I_a$,
of gauge group SU(2) to study the topological content of gauge field
configurations in Monte Carlo generated ensembles.
Since the fermion is in the real representation of the gauge group, the
spectrum of the hermitian Wilson Dirac operator, $\ham_L(m)$, is doubly
degenerate. Therefore, the index of the associated chiral Dirac
operator can only be even valued. This is to be expected. Otherwise,
we would have to explain away unphysical expectation values of an odd
number of fermionic observables in an otherwise well defined theory.
However, it is possible to obtain any even value for the index. The issue
one has to address is if all possible even values are realized for the
index in an ensemble of SU(2) gauge field backgrounds or if only
multiples of four are observed. If only multiples of four are observed
then one would conclude that the gauge field background behaves as if
it were made up of classical instantons with small fluctuations, and we
would not be able to explain the N degenerate vacua by topological
means. On the other hand, if we observe any even value for the index
that are not just multiples of four, the background gauge fields
cannot be thought of as being made up of classical instantons and we
would have evidence for N degenerate vacua arising out of a topological
mechanism.

We considered fifty configurations, each, in the two SU(2) ensembles already
used to compute the topological susceptibility from the index $I_f$ in
the fundamental representation, shown in Fig.~\ref{fig:chi}~\cite{su2_adj}.
We have doubled this number by including the parity transformed partner of
every configuration since this symmetrize's the distribution of the indices.
The computation of the index of the overlap chiral Dirac operator in the
adjoint representation was done exactly as for the fundamental
representation. We found configurations for which $I_a=4I_f$, {\it
i.e.,} where the topological charge as given by the fundamental and
adjoint indices agrees. The corresponding zero modes were compatible
with representing the same topological objects. But we found also
many configurations in both ensembles that do not satisfy the
relation $I_a=4I_f$, and we found several configurations where $I_a$ is
not a multiple of four.  The occurrence of a significant number of
configurations with values of $I_a$ that are not multiples of four is
taken as evidence for the existence of gauge field configurations with
fractional topological charge in our ensemble since $Q={I_a\over 4}$
is the continuum relation between the topological charge and the index
of the Dirac operator in the adjoint representation. 

\begin{figure}
\epsfxsize=3.0in
\centerline{\epsfbox[10 80 586 556]{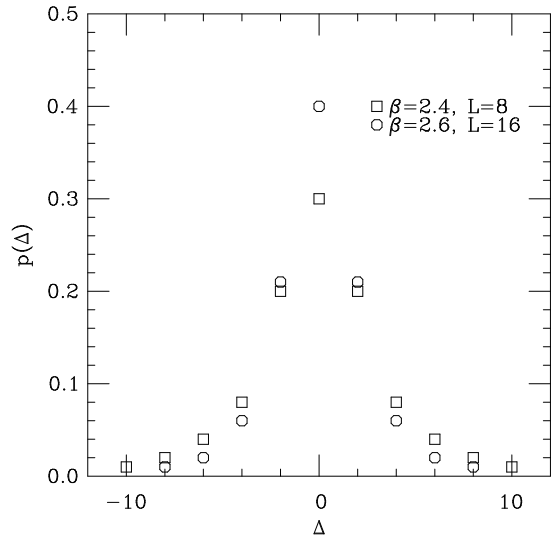}}
\caption{$p(\Delta)$ versus $\Delta$ for the two ensembles where
$\Delta=I_a-4I_f$.}
\label{fig:delta}
\vskip -3mm
\end{figure}

Having provided some evidence for the existence of fractional
topological charge on the lattice at finite lattice spacing, we now
address the question of whether these are pure lattice artifacts. For
this purpose, we define the quantity $\Delta=I_a-4I_f$ for each
configuration in both ensembles.  Note that $\Delta$ takes on only
even values. The probability of finding a certain value of $\Delta$,
$p(\Delta)$, is plotted for the two ensembles in
Fig.~\ref{fig:delta}. We find that $p(\Delta)$ for $|\Delta| > 2$
decreases as one goes toward the continuum limit at a fixed physical
volume. However, $p(\pm 2)$ does not decrease indicating that it
might remain finite in the continuum limit.

In summary we have presented some preliminary evidence for
fractional topological charge on the lattice. By studying two
ensembles with different lattice spacings we have argued that there is
a reasonable indication that this is not a lattice artifact.  If this
result survives the continuum limit it provides a topological basis
for N degenerate vacua in supersymmetric SU(N) gauge theories.

\section*{Acknowledgements}

This research was supported by DOE contracts 
DE-FG05-85ER250000 and DE-FG05-96ER40979.
Computations were performed on the QCDSP and CM-2 at SCRI.


\end{document}